\documentclass[fleqn,12pt,twoside]{article}
\usepackage{espcrcproc}
\usepackage{epsfig} 
\usepackage{wrapfig}

\usepackage{graphicx}
\usepackage[figuresright]{rotating}

\def\Journal#1#2#3#4{{#1} {\bf #2}, #3 (#4)}

\def\PLB{{\em Phys. Lett.}  B}
\def\PRL{\em Phys. Rev. Lett.~}
\def\PRD{{\em Phys. Rev.} D}

\newcommand{\pom}{I\!\!P}
\newcommand{\reg}{I\!\!R}
\newcommand{\alphapom}{\alpha_{_{I\!\!P}}}
\newcommand{\alphareg}{\alpha_{_{\rm I\!R}}}
\newcommand{\xpom}{x_{\pom}}
\newcommand{\xPom}{\xpom}
\newcommand{\xbj}{$x_{Bj}$} 
\newcommand{\qsq}{\mbox{$Q^2$}}
\newcommand{\bet} {$\beta$}

\newcommand{\ttt} {$t$}

\newcommand{\fdthree} {$F_2^{D(3)}$}
\newcommand{\fdthreef} {$F_2^{D(3)}  (Q^2, x_{I\!\!P} , \beta )$}
\newcommand{\fdtwo} {$F_2^{D(2)} (Q^2,\beta)$}

\title{Experimental review of diffractive phenomena}

\author{L.~Favart\address{I.I.H.E., CP-230\\
        Universit\'e Libre de Bruxelles, \\
        1050 Brussels\\
        Belgium\\
        lfavart@ulb.ac.be
        }%
       }
\begin{document}

\maketitle

\begin{abstract}
A review is given of the measurements of the hard
diffractive interactions in recent years from two high-energy colliders,
the HERA $ep$ collider and the Tevatron $p\bar{p}$ collider. The 
structure of the diffractive exchange in terms of partons, the
factorisation properties and the ratio of diffractive to non-diffractive
cross sections are discussed.
\end{abstract}

\section{INTRODUCTION}

The diffractive interaction is a feature of hadron-hadron scattering
at high energy corresponding to a $t$-channel exchange of the vacuum
quantum numbers and a small momentum transfer.
It has been has been described, in the past, in the framework of Regge 
theory, where the exchange was interpreted as the Pomeron ($\pom$) trajectory, 
characterized by a weak energy dependence, in particular, with respect to
the fast decrease of the total cross section at smaller energy due to
Reggeon ($\reg$) exchange:
$\sigma_{Tot}=B\ W^{2(\alphareg -1)} + \ A \ W^{2(\alphapom -1)}$,
where $W$ is the center of mass energy, $A$ and $B$ normalisation
factors and typically $\alphareg=0.55$ and $\alphapom=1.08$.
\\

\begin{wrapfigure}[13]{L}{0.70\linewidth}
 \vspace*{0.5cm}
 \begin{picture}(60,80) 
  \put(0,-5){\epsfig{figure=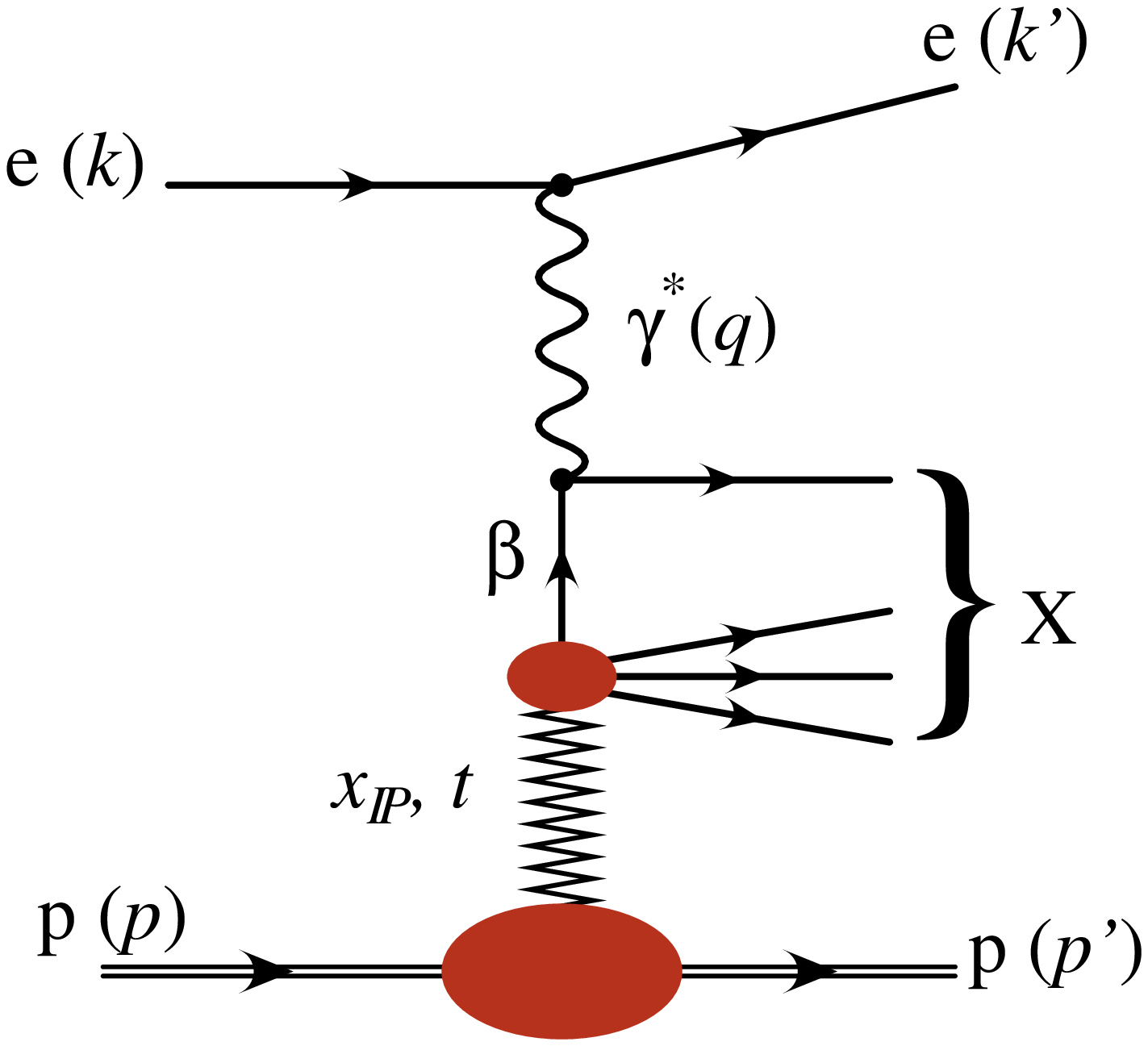,width=0.30\textwidth}} 
  \put(150,-5){\epsfig{figure=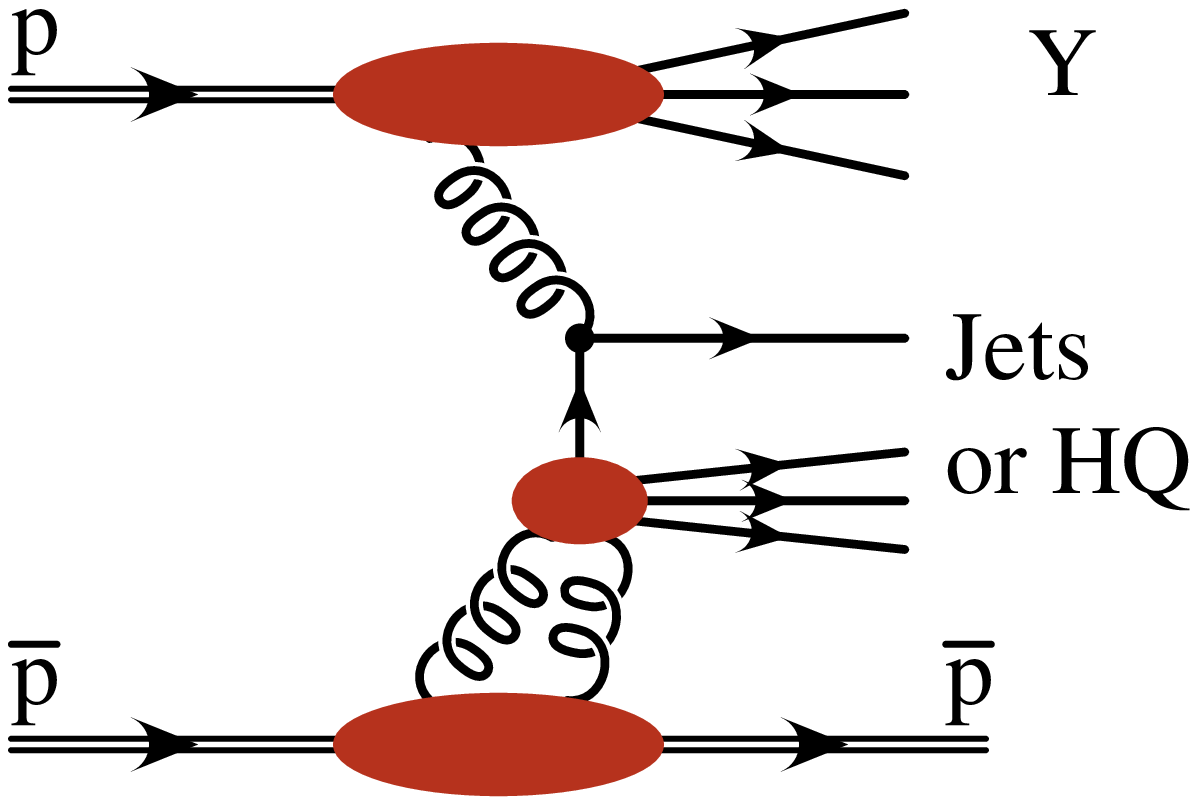,width=0.35\textwidth}} 
  \put(35,110){a) HERA}
  \put(180,110){b) Tevatron}
 \end{picture}
 \caption{Basic diagrams for diffraction in presence of a hard scale at
 HERA and at Tevatron}
 \label{fig:diag}
\end{wrapfigure}

Diffraction, directly related to the total cross section through the
optical theorem, has two distinguishing features. 
First, hadron emission from the exchange itself being suppressed
by its colourless nature, the two diffractively dissociated
systems are separated in rapidity space, forming
a large rapidity gap (LRG).
Secondly, the diffractive events are observed with a small momentum 
transfer in both the transverse and longitudinal coordinates.
The four-momentum of the exchange, $t$,
and the longitudinal
momentum fraction of the exchange, $\xpom$, are both small;
$|t|$ is typically less than the square of the nucleon
mass and $\xpom$ is smaller than 0.05.
\\

The high energies of the HERA $ep$ collider and of the Tevatron
$p\bar{p}$ collider allow us for the first time to study diffraction 
in term of perturbative QCD (pQCD), i.e.~in the presence of a hard
scale, which is called hard diffraction. 
At HERA the diffractive interaction takes place between the
hadronic behaviour of the exchange virtual photon and the proton
(see Fig.~\ref{fig:diag}a), and 
between the two protons at Tevatron (see Fig.~\ref{fig:diag}b). 
The possible hard scales are large $Q^2$, 
the negative of the four-momentum squared of the exchanged
virtual photon, large transverse energy, $E_T$, in jet production,
heavy-quark mass or large $t$.
The hard scale presence of large $Q^2$ at HERA and large  $E_T$ at Tevatron,
in particular,
gives the possibility to probe the partonic structure of the diffractive
exchange. As in inclusive inelastic scattering (DIS)~\cite{Vladimir}, 
the presence of a hard scale in the process is characterized by an 
important positive increase of the energy dependence behaviour. 
\\

This article concentrates on two topics, the factorisation properties
and the ratio between diffractive and non-diffractive structure functions. 
Both give insight into the understanding of the nature of diffraction in
terms of partons. 
Before reviewing recent measurement, a short
discussion on the parton densities of the diffractive exchange and their 
relation to the diffractive cross section is given.
\\

Although it will not be covered here, the study of diffraction has
several other topics of interest in top of the diffractive exchange measurement
in terms of partons. It gives an access to the nucleon structure
at low \xbj and to the parton correlations through the generalized
parton distributions (GPDs) approach. It allows to test the region of validity
of the different asymptotic dynamical approaches of QCD that are DGLAP 
and BFKL. Finally, the alternative approach of colour Dipole models 
allows us to study the transition between non-perturbative and perturbative
regimes and to test the presence of a gluon density saturation in the
proton. 

\section{PARTONIC STRUCTURE OF DIFFRACTIVE EXCHANGES AND FACTORISATION
PROPERTIES}

The inclusive diffractive cross section at HERA, $e p \rightarrow e X p$,
can be defined with the help of four kinematic variables
conveniently chosen as  \qsq , $\xpom$ , \bet\ and \ttt, where
$\xpom$ \ and \bet\ are defined as

$$ 
   \xpom \simeq \frac{Q^2 + M_X^2}{Q^2 + W^2 },
   \qquad \qquad 
    \beta \simeq \frac{Q^2 }{Q^2 + M_X^2} ; \label{eq:kin}
$$ 
$M_X$ being the invariant mass of the X system.
$\xpom$ can be interpreted as the fraction of the
proton momentum carried by the exchanged
Pomeron and \bet\  is the fraction of the exchanged momentum carried by
the quark struck by the photon, or in other terms, the fraction of the
exchanged momentum reaching the photon.
These variables are related to the Bjorken \xbj\ scaling variable
by the relation \xbj$= \beta \cdot \xpom$.
The presence of the hard scale, $Q^2$, ensures that the virtual photon
is point-like and that the photon probes the partonic 
structure of the diffractive exchange (Fig.~\ref{fig:diag}a),
in analogy with the inclusive DIS processes.
\\

{\bf Factorisation Properties in hard Diffraction.}\\
For hard QCD processes in general, like high-$E_T$ jet production or
DIS,
the cross section can be factorized into two terms; the parton
density and the hard parton-parton cross section. In the case of
DIS, the cross section can be written as
$$
\sigma = \sum_i {f_i(\xi, \mu^2) \; \hat{\sigma}_{i\gamma}(\xi, \mu^2)}
$$
where $i$ runs over all parton types, $f_i$ is the parton density
function for the $i$-th parton with longitudinal momentum fraction
$\xi$, which is probed at the factorisation scale $\mu$.
$\hat{\sigma}_{i\gamma}$
denotes the cross section for the interaction of
the $i$-th parton and the virtual photon.
Such an expression, often referred to as the QCD factorisation theorem,
is well
supported by the data. If the theorem holds,
only one parton per hadron is coupled to the hard scattering vertex.
\\

The theorem is proven to be applicable also for 
hard inclusive diffraction~\cite{collins_fact} in $ep$ collisions 
at large \qsq\, namely
$$
\frac{d\sigma(x,Q^2,\xPom,t)}{d\xPom dt} = 
     \sum_i\int^{\xPom}_{x} dz \;
      \hat{\sigma}_{i\gamma}(z, Q^2, \xPom) \;
             f_i^D(z, Q^2, \xPom, t)
$$
where $z$ is the longitudinal momentum fraction of the parton
in the proton, $\sigma_{i\gamma}$ is again the hard scattering
parton-photon cross section for hard diffraction and $f_i^D$
is the diffractive parton density for the $i$-th parton.
$f_i^D$ can be regarded as the parton density of the diffractive
exchange occuring at a given $(\xPom, t)$.
If such a theorem holds, $f_i^D$ should be universal for all hard
processes,
e.g.\ inclusive diffraction, jet or heavy-quark production etc \ldots
\\

Experimentally, the \ttt\ variable is usually not measured and is
integrated over.
In analogy with non-diffractive DIS scattering, the measured cross
section
is expressed in the form of a three-fold diffractive structure function
\fdthreef\ (neglecting the longitudinal contribution), 
$$
  \frac { {\rm d}^3 \sigma \ (e p \rightarrow e X p) }
 { {\rm d}Q^2 \ {\rm d}\xpom \ {\rm d}\beta}
        = \frac {4 \pi \alpha^2} {\beta Q^4}
            \ (1 - y + \frac {y^2}{2} )
            \ F_2^{D(3)} (Q^2, \xpom , \beta) ,
                                            \label{eq:fdthreefull}
$$
where $y$ is the usual scaling variable, with
$  y \simeq W^2 / s$.
\\

\begin{figure}[htb]
 \vspace*{3.2cm}
 \begin{picture}(60,70) 
  \put(0,-50){\epsfig{file=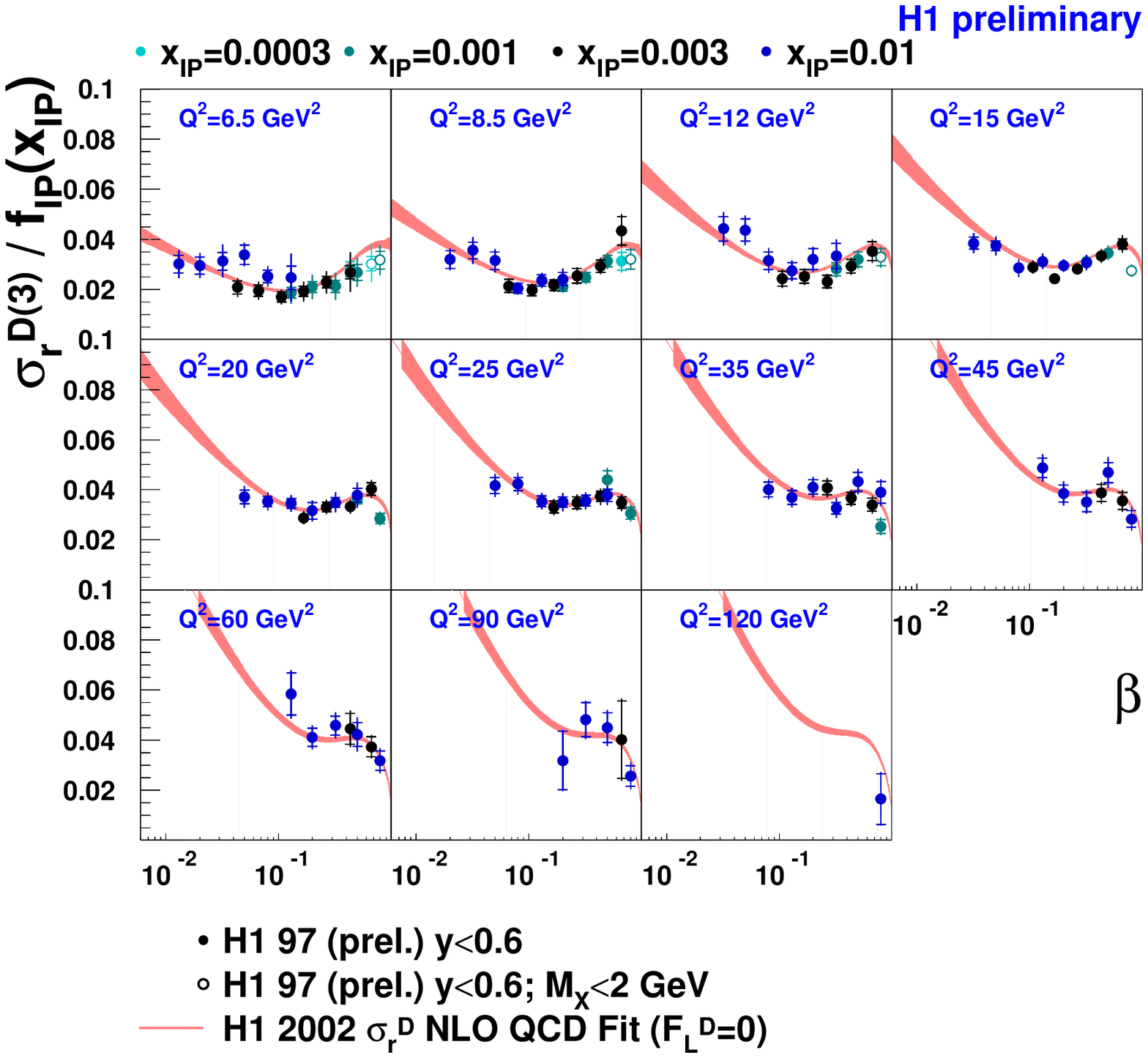,width=0.52\linewidth,clip=}}
  \put(217,-15){\epsfig{file=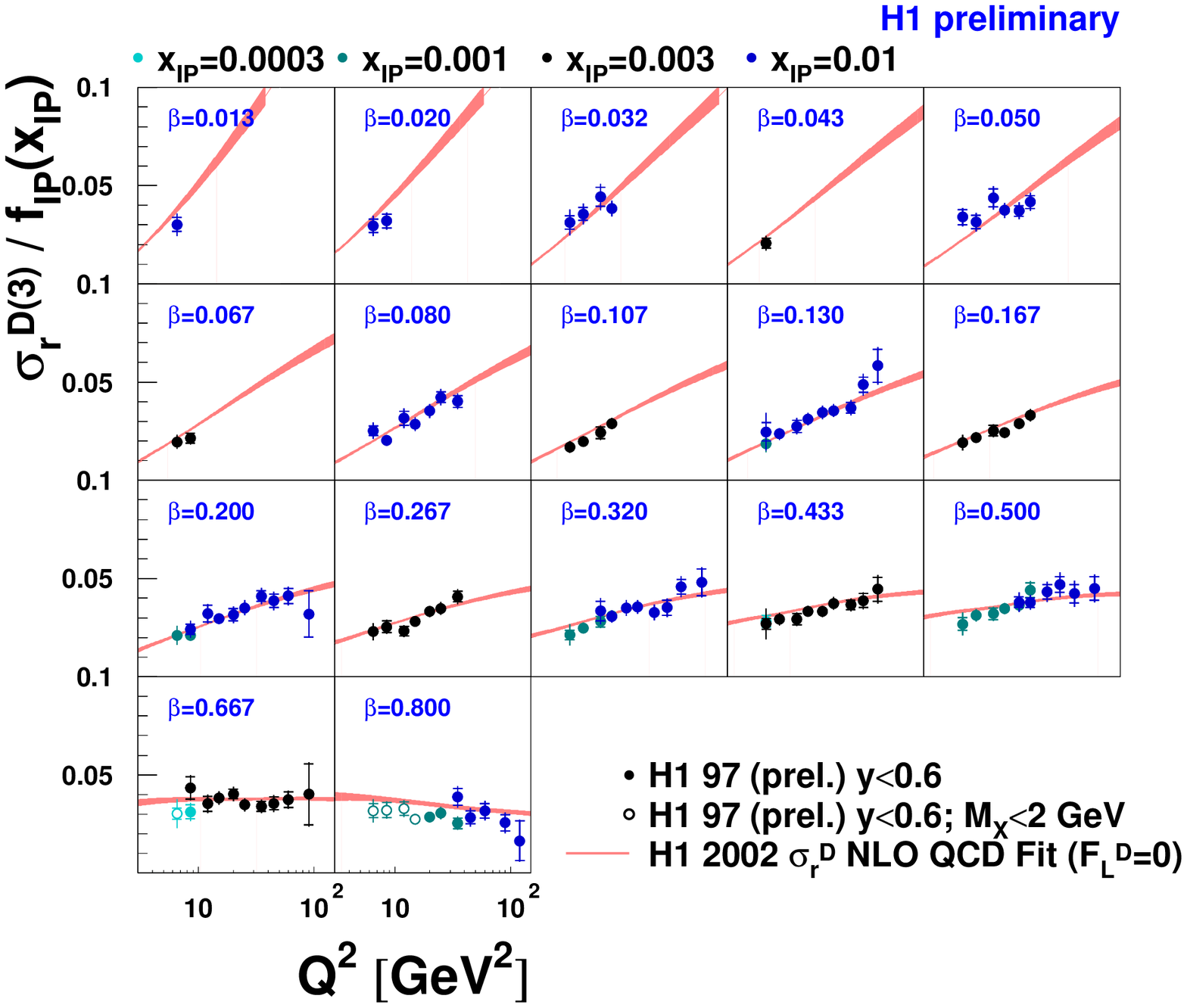,width=0.52\linewidth}}
 \end{picture}
 \caption{The \fdtwo\ diffractive structure function measurement 
for different values of $\xpom$ as a function
of \qsq\ in bins of \bet\ (left plot) and as a function of  \bet\ in
bins of \qsq\ (right plot). 
}
 \vspace*{-0.5cm}
 \label{fig:f2d}
\end{figure}

Conveniently, the Regge factorisation is applied where \fdthree \ is 
written in the form
$$F_2^{D(3)} (Q^2, \xpom , \beta ) =
     f_{\pom/p} (\xpom ) \cdot F_2^D (Q^2, \beta),$$
assuming that the $\pom$ flux $ f_{\pom/p} (\xpom)$ is independent
of the  $\pom$ structure \fdtwo\ ,
by analogy with the hadron structure functions, \bet\ playing the role
of Bjorken \xbj. The  $\pom$ flux is parametrised in a Regge inspired
form $f_{\pom/p} = e^{bt}/\xpom^{2\alphapom(t) - 1}$, where the value of
the Pomeron intercept $\alphapom(0)=1.173$ is used.
Including the Reggeon ($\reg$) trajectory 
in addition to the Pomeron,
a good description of the data is obtained throughout
the measured kinematic domain by H1~\cite{h1f2d97} and ZEUS \cite{zeuscharm}
experiments.
\\

\begin{wrapfigure}[21]{L}{0.50\linewidth}
 \vspace{-0.8cm}
  \epsfig{file=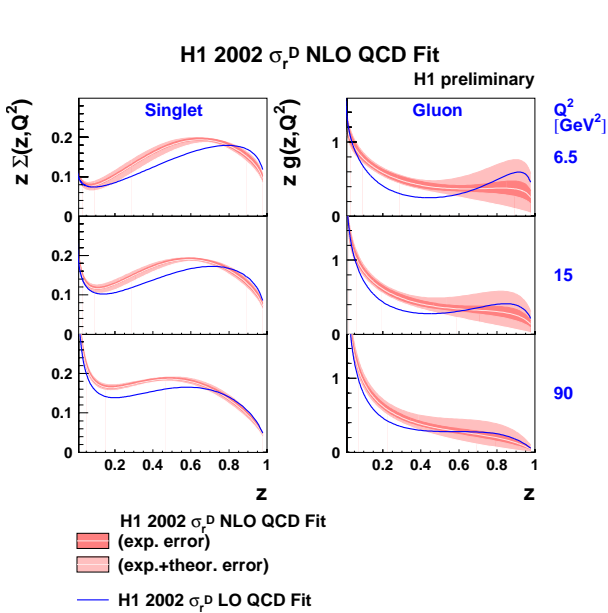,width=0.50\textwidth}
 \caption{Quark (singlet) and gluon densities in the $\pom$ extracted 
  from the QCD fit of \fdtwo as a function of $z$.}
\label{fig:gluond}
\end{wrapfigure}
The measured points and 
the fitted structure function \fdtwo\ from H1~\cite{h1f2d97} 
is shown in Fig.~\ref{fig:f2d},
by the ratio of the cross section to the $\pom$ flux, for different values
of $\xpom$, 
as a function of \bet\ for fixed \qsq\ on the left plot and as a
function of \qsq\ for fixed \bet\ on the right plot.
The left plot shows that the Regge factorisation ansatz holds within 
the present precision of the measurement as the \fdtwo\ measurement is not
sensitive to the $\xpom$ value. The \qsq\ dependence shown on the right
plot exhibits a more important scaling violation than in the $F_2$
structure function measured in non-diffractive deep inelastic
scattering~\cite{Vladimir}. This indicates that the exchanged object 
in diffraction has an important gluon contain.
\\

 By analogy to the QCD evolution of the proton structure function $F_2$,
one can attempt to extract the partonic structure of the Pomeron
from the \qsq\ evolution of \fdtwo. Starting the QCD evolution at
$Q^2_0=6.5 \rm\ GeV^2$, H1 has extracted the partonic 
distributions \cite{nlopaper} as shown in Fig.~\ref{fig:gluond}
separately for the gluon
and the singlet quark components as a function of $z$, the Pomeron
momentum fraction carried by the parton entering the hard interaction.
This distribution shows the dominance of gluons up to the high $z$
values in the Pomeron partonic structure (75$\pm$15\% after
integration in $z$). 
\\

\begin{figure}[htb] \unitlength 1mm
   \epsfig{file=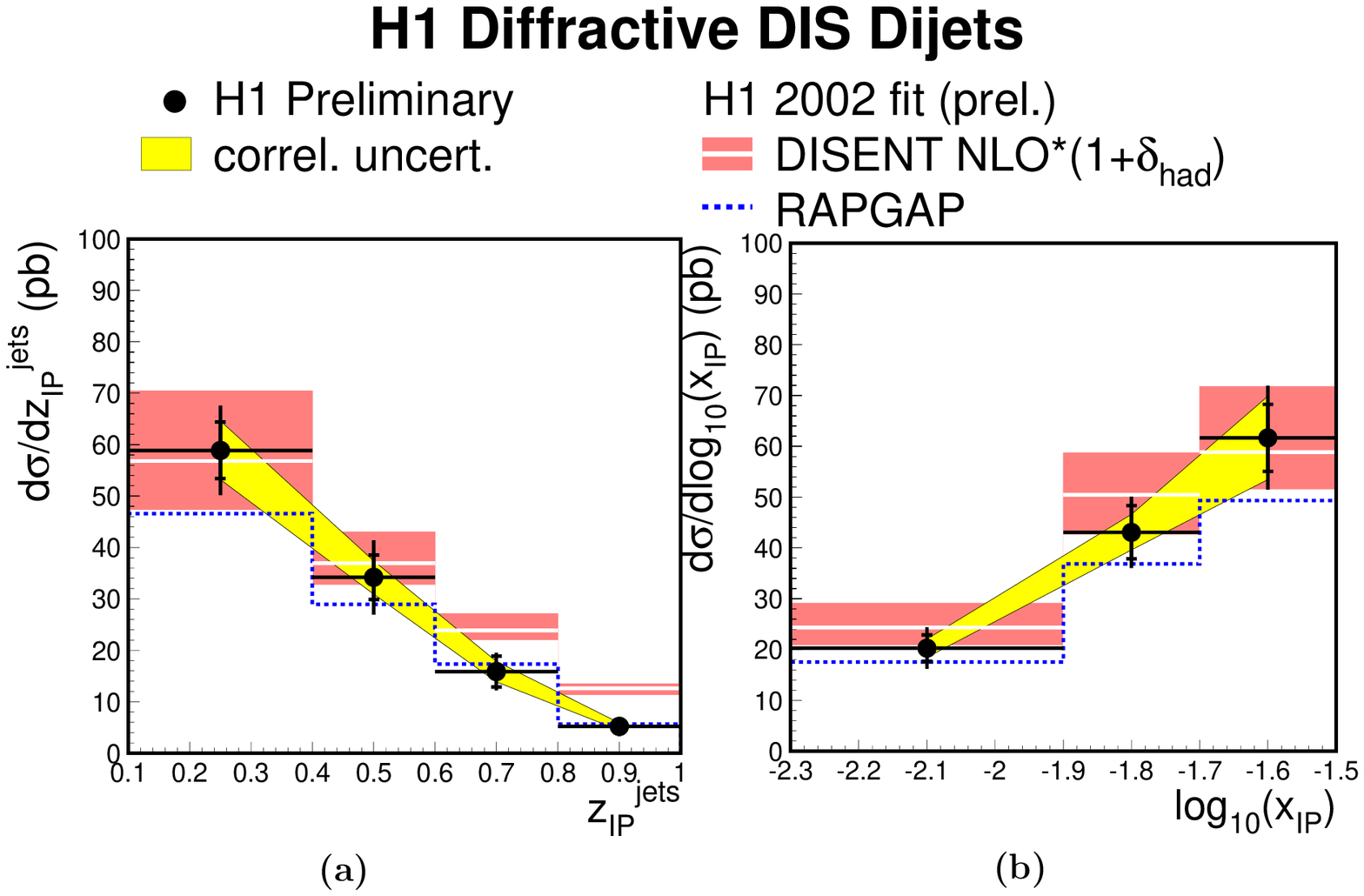,width=0.60\textwidth}
   \epsfig{file=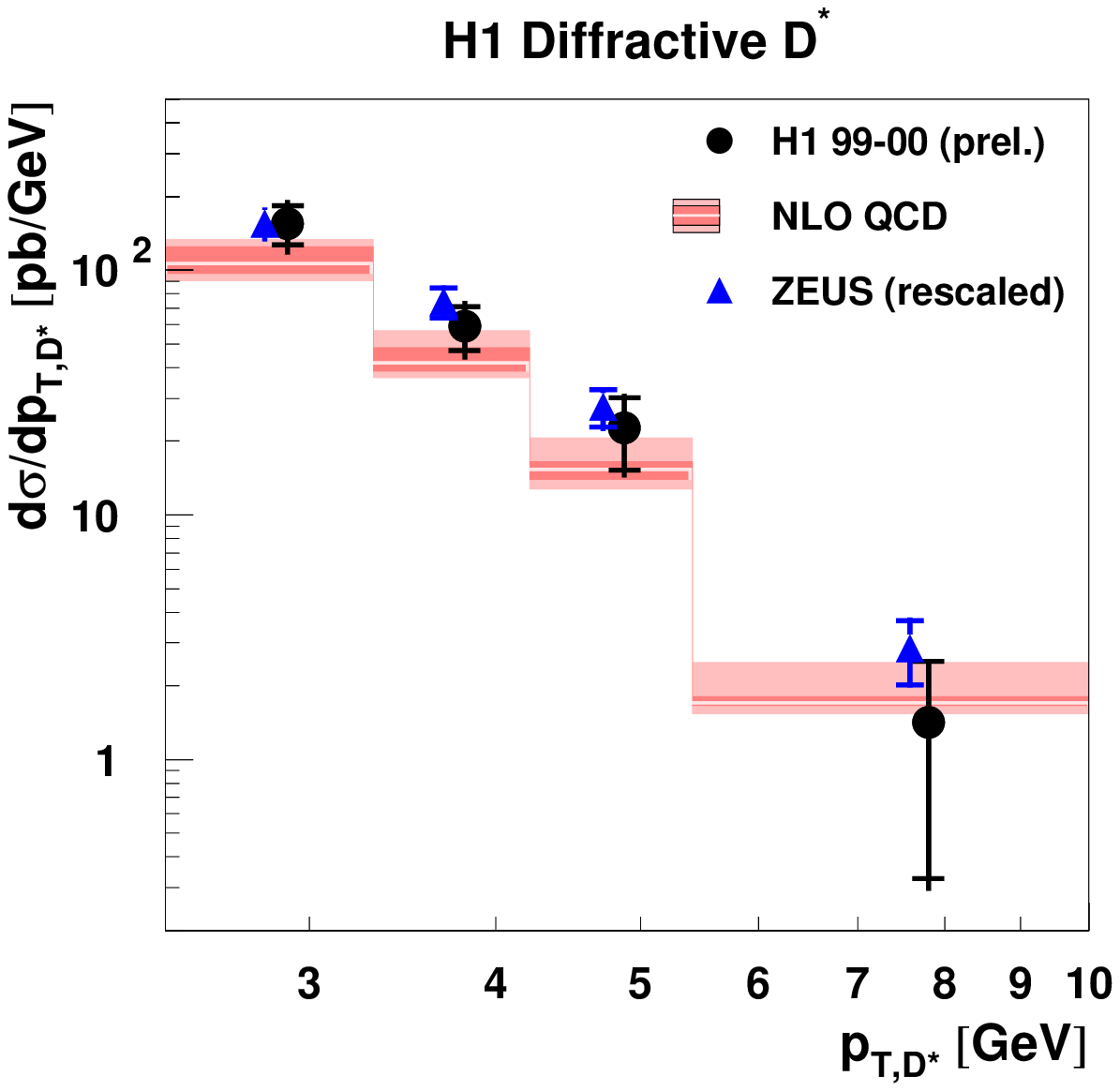,width=0.40\textwidth}
 \caption{{\bf left}: H1 measurement of diffractive dijet cross section 
 in electroproduction as a 
 function of $z_{\pom}^{jets}$, an estimator for the parton momentum fraction
 of the diffractive exchange entering the hard sub-process, and as a
 function of $\xpom$ as measured by H1 and ZEUS.
 {\bf right}: Diffractive $D^*$ meson cross sections in
electroproduction differential in $p^*_{T,D^*}$.
 }
\label{fig:dijet}
\end{figure}

{\bf Dijet and charm productions in diffractive electroproduction}\\
To test QCD factorisation for diffractive dijet production in
electroproduction regime ($Q^2 >> 1\rm\ GeV^2$), the
H1 dijet cross section~\cite{h1dijet} in the kinematic range 
$Q^2>4 \rm\ GeV^2$ and $\xpom<0.03$ is compared to the NLO
QCD prediction in Fig.~\ref{fig:dijet}, using the extracted 
diffractive parton densities. The cross
sections were corrected to asymmetric cuts on the jet transverse
momentum $p_{T,1(2)}>5(4) \rm\ GeV$, to facilitate comparisons with
NLO calculations.
The inner error band of the NLO
calculations represents the renormalization scale uncertainty, whereas
the outer band includes the uncertainty in the harmonization
corrections. Within the uncertainties, the data are well described in
both shape and normalisation by the NLO calculations, in agreement with 
QCD factorisation.
\\

The  $D^*$ meson production measurements in
diffractive electroproduction were published by H1~\cite{h1charm} and
ZEUS~\cite{zeuscharm} for the kinematic range $Q^2>2 \rm\ GeV^2$,
$\xpom<0.03$  
and $p^*_{T,D^*}> 2 \rm\ GeV$, where the latter
variable corresponds to the transverse momentum of the $D^*$ meson in
the photon-proton centre-of-mass frame.
NLO QCD calculation were performed interfacing the H1 diffractive parton
distributions.
The renormalization and factorisation scales were set to
$\mu^2=Q^2+4m_c^2$. Further parameter values are a charm quark mass
of $m_c=1.5 \rm\ GeV$, a $c\rightarrow D^*$ harmonization fraction of
$f(c\rightarrow D^*)=0.233$ and $\epsilon=0.078$ for the used Peterson
fragmentation function.
A comparison of the calculations with the $D^*$ H1 and ZEUS data is shown in
Fig.~\ref{fig:dijet}-right. The inner error band of the NLO calculation
represents the renormalization scale uncertainty, whereas the outer
error band includes variations of $m_c$ and $\epsilon$. Within the
uncertainties, the data are well described in both shape and
normalisation by the NLO calculations, supporting the idea of QCD 
factorisation.

\section{HARD DIFFRACTION AT THE TEVATRON}


At the Tevatron, the diffractive process occurs in
various combinations of the large rapidity gap configuration
since the both incoming protons can emit
the diffractive exchange. The partonic content of the exchange
is mainly studied using single-diffractive
processes, where the colour-octet parton emitted from one proton
undergoes a hard scatter with the diffractive exchange emitted
from the other proton; the partons in the exchange
are probed by the colour-octet parton replacing the virtual photon at
HERA.
$$
 \epsfig{file=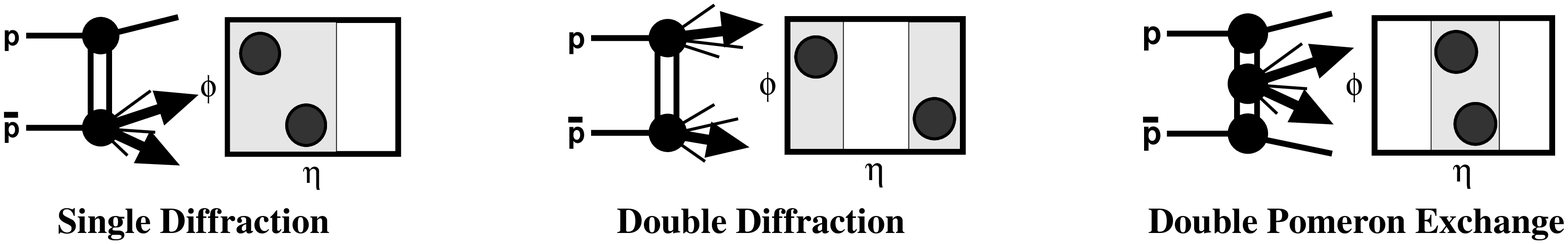,width=0.8\textwidth,clip=}\\ 
$$

Various hard processes have been studied in Tevatron Run I data, such as
$W$ production~\cite{Tevatron-SD-W} which is sensitive
to the quark content of the exchange, and
dijet\cite{Tevatron-SD-dijet-1,Tevatron-SD-dijet-2}
and $b$-quark production\cite{Tevatron-SD-HQ},
which are more sensitive to the gluonic content.
\\

\begin{figure}[htb]
 \vspace*{-1.4cm}
 \epsfig{figure=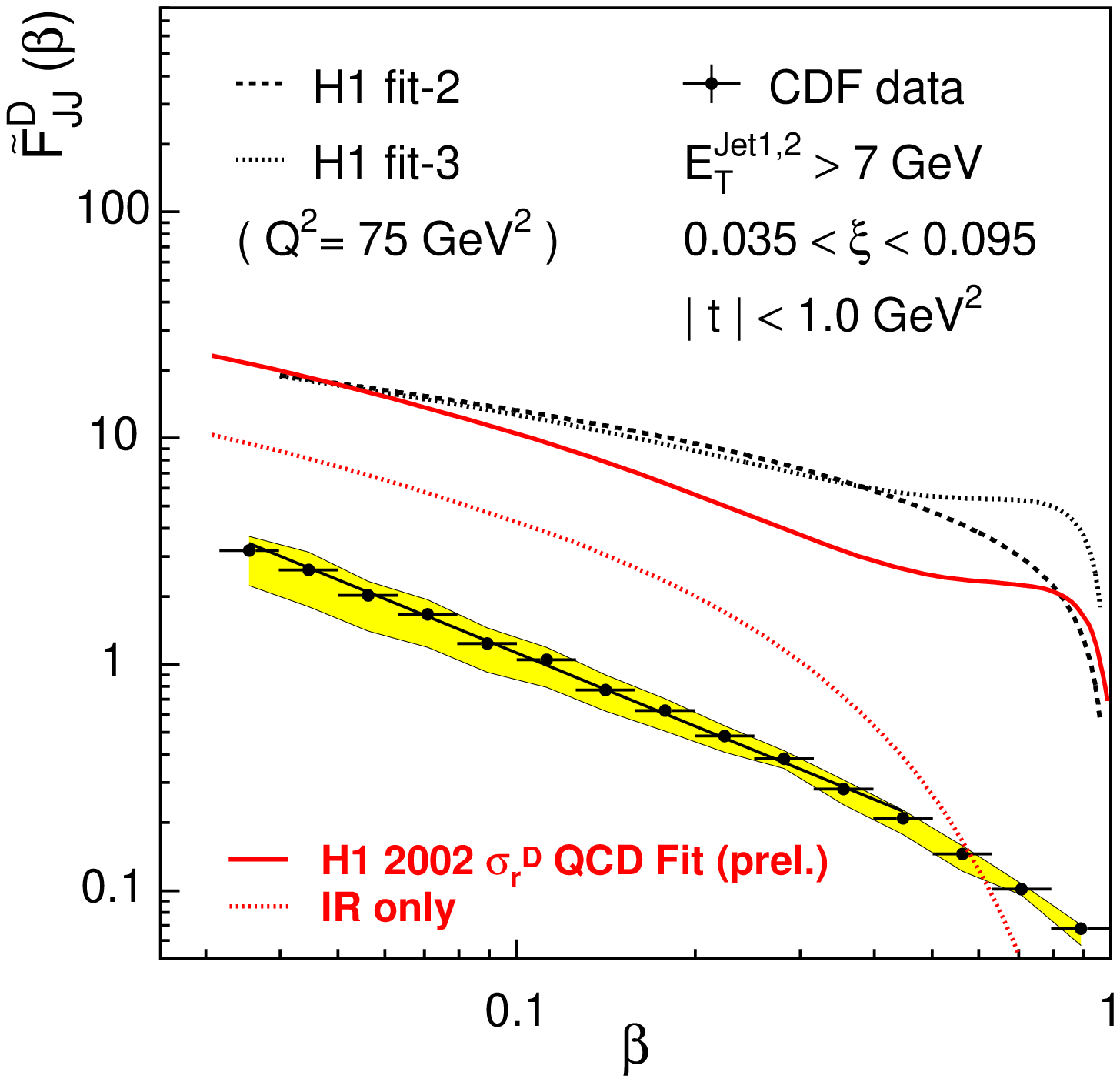,width=0.52\textwidth}
 \epsfig{figure=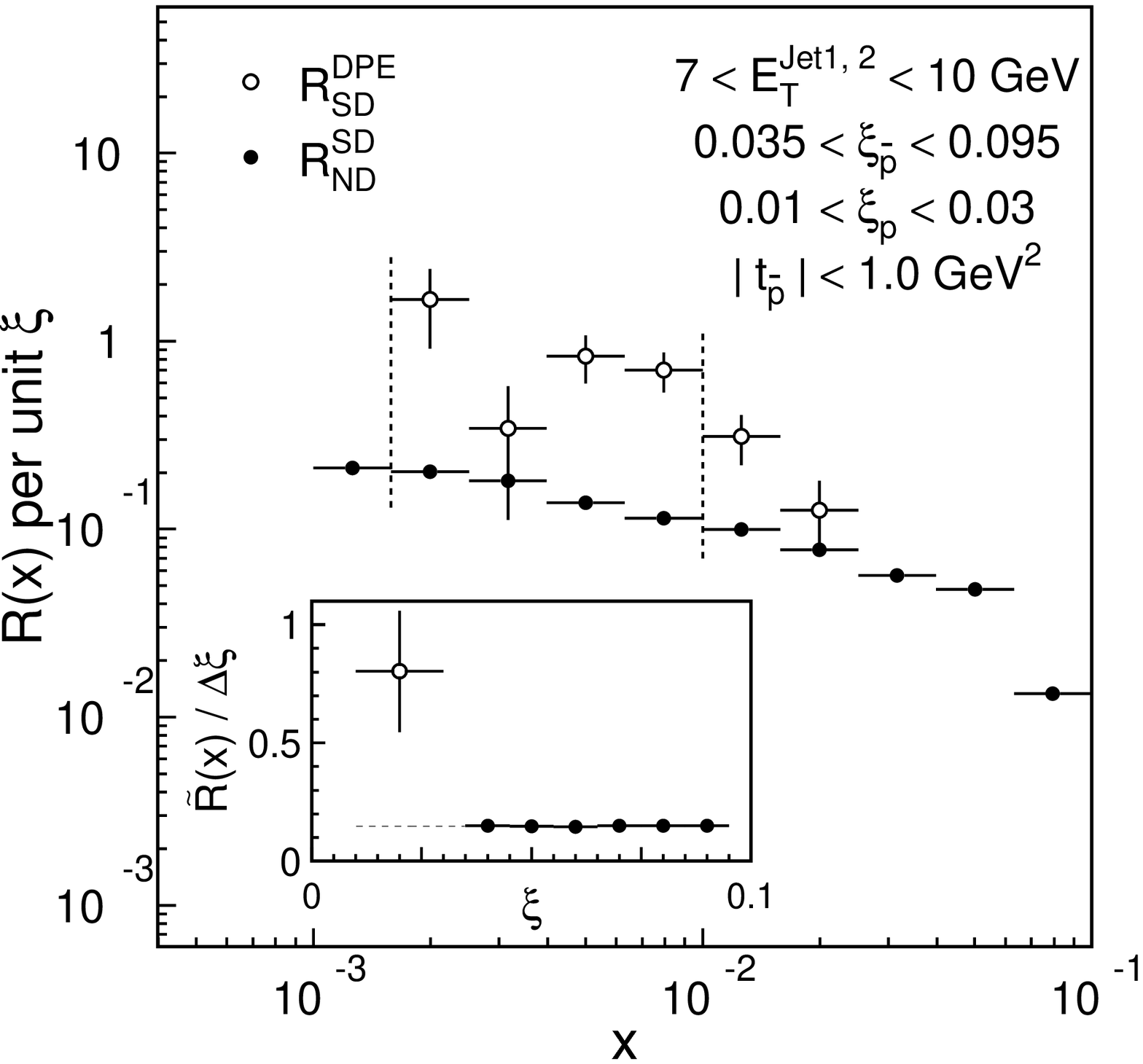,width=0.47\textwidth}
 \vspace*{-1.0cm}
\caption[]{{\bf left }: measured dijet cross section in single-diffractive
processes in comparison with the prediction using the diffractive PDF's 
obtained by H1. {\bf right }:  measured dijet cross section in
single-diffractive processes compared to double-diffractive
processes by CDF. 
}
\label{fig:tevdijet-h1fit}
\end{figure}

{\bf Hard diffraction at the Tevatron from Run I data
and the factorisation breaking}\\
One of the most striking features of the hard diffractive process
measured at the Tevatron is a large suppression of the cross section
with respect to the prediction based on the diffractive
parton densities obtained from the HERA $F_2^D$ data.
Figure~\ref{fig:tevdijet-h1fit}-left shows the comparison of the dijet
cross section in the single-diffractive process measured by CDF 
at the Tevatron~\cite{Tevatron-SD-dijet-1} to the
prediction using the diffractive parton densities discussed in
the previous section. Although the prediction reproduces
the shape of the data in the low-$\beta$ region, the magnitude of the
cross section is smaller by about a factor 5 to 10.
A similar degree of suppression was observed in the other hard
diffractive processes~\cite{Tevatron-SD-W,Tevatron-SD-dijet-2}.
This indicates a strong factorisation breaking
between HERA and the Tevatron: the diffractive parton densities
are not universal between these two environments.
Additionally, CDF has measured the ratio of single-diffractive over
double-diffractive processes (shown on in the right plot of
Fig.~\ref{fig:tevdijet-h1fit}) to be of 0.19 $\pm$ 0.07. This indicates 
that the formation of a second gap is not (or only slightly) suppressed.  
\\

The reason for the breaking is not yet clearly known.
It is usually attributed to re-scattering between spectator partons in the 
two beam remnants where one or more colour-octet partons are exchanged,
which destroys the already formed colour-singlet state.
\\

{\bf Factorisation test with hard photoproduced diffraction at
HERA}\\
QCD factorisation can be tested within HERA
looking at the diffractive dijet photoproduction ($Q^2 \sim 0$),
where the hard scale is provided by the $E_T$ of the jets.
Factorisation is expected not to hold in photoproduction
events, where the resolved process has a photon remnant, allowing
re-scattering.
On the other hand, the direct process does not have
a beam remnant and the suppression of diffractive events is expected
to be much smaller than in the resolved process.
\\

\begin{figure}[tb]
 \vspace*{5.3cm}
 \begin{picture}(60,70) 
  \put(0,-30){\epsfig{figure=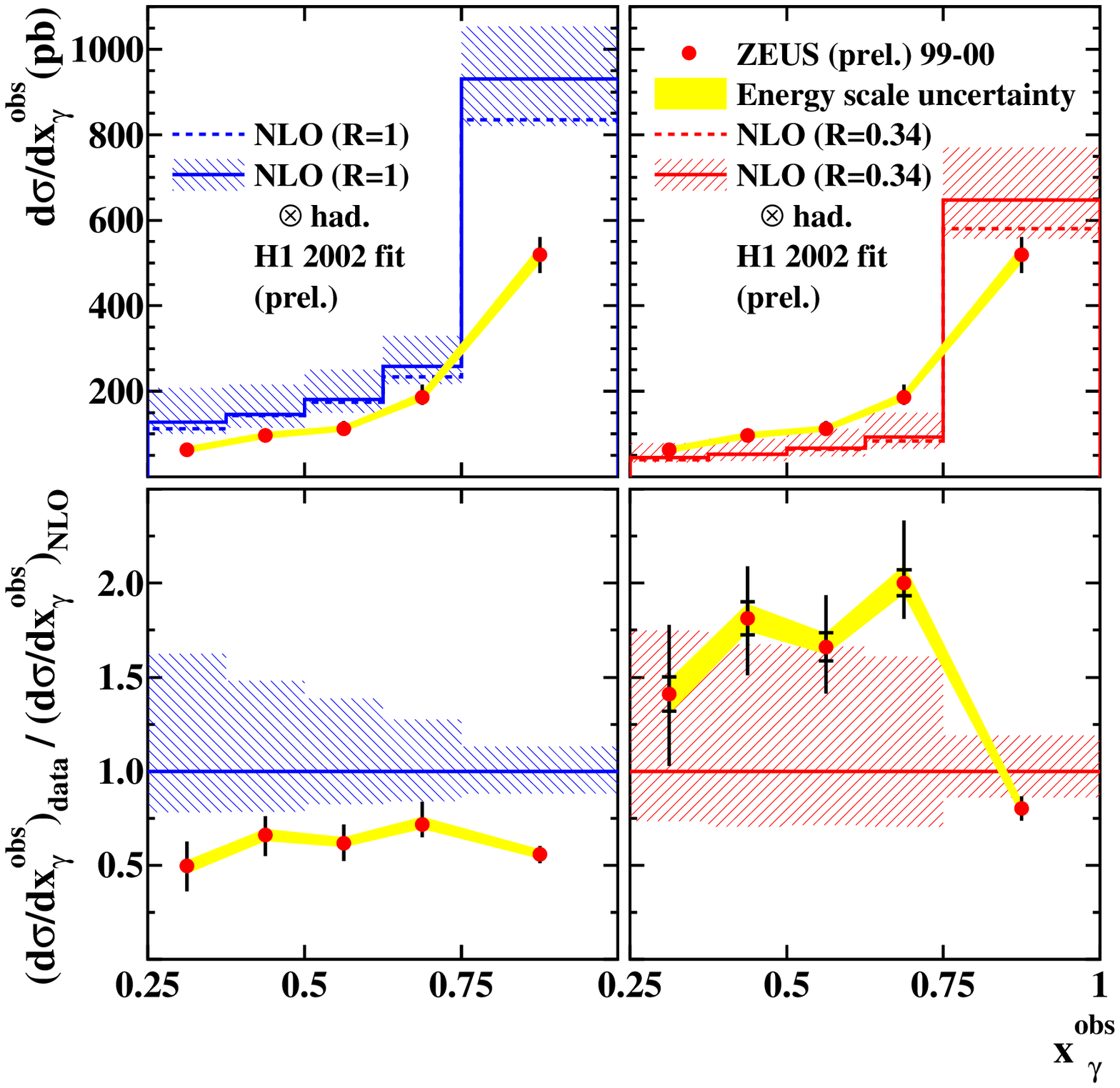,width=0.60\textwidth}}
  \put(280,40){\epsfig{file=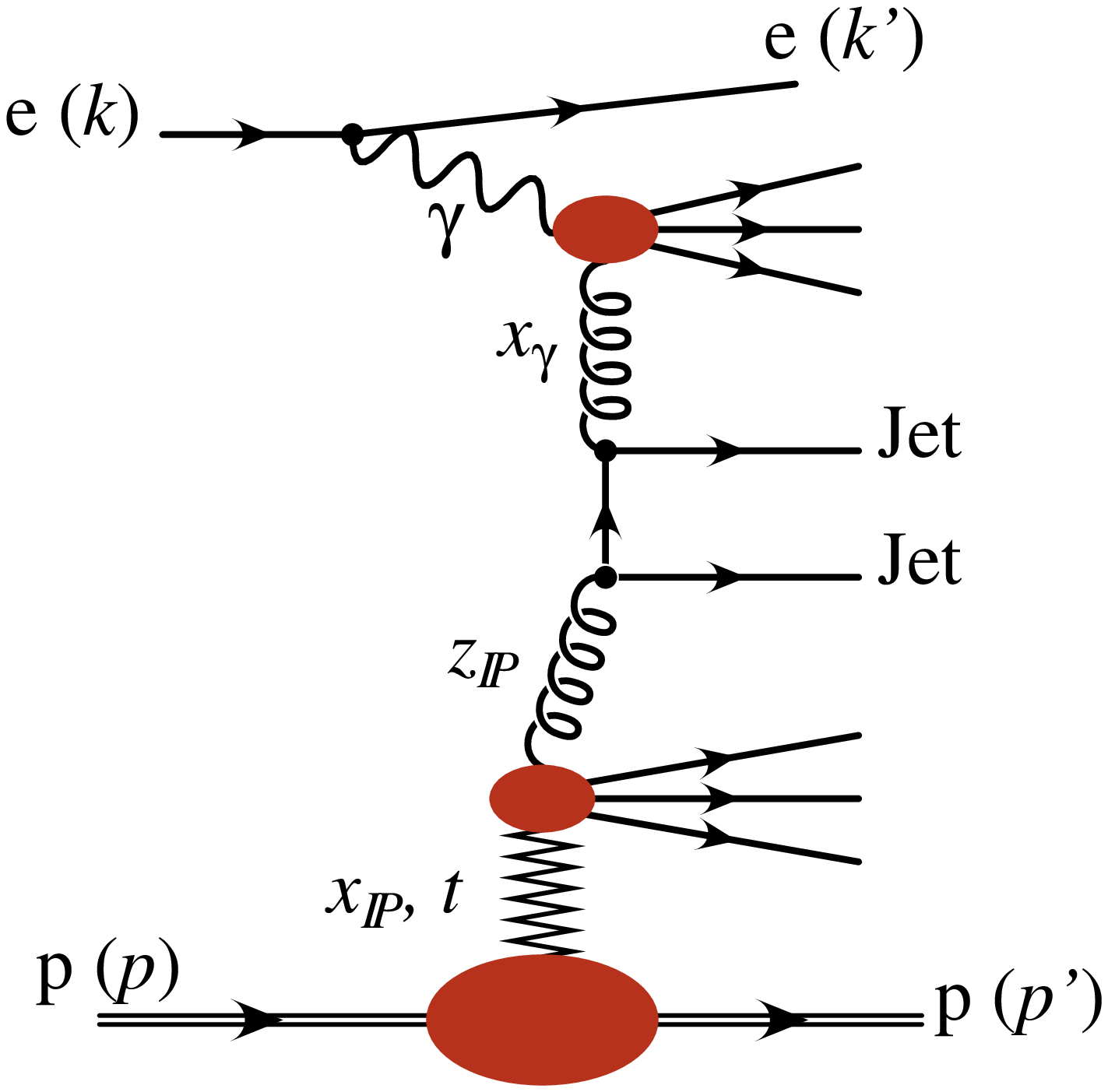,width=0.40\textwidth}}
 \end{picture}
\caption[]{{\bf left}: Dijet cross section in photoproduction at HERA 
measured by ZEUS as a function of $x_\gamma^{\rm jets}$. 
{\bf right}: Diagram for dijet in photoproduction at HERA.
}
\label{fig:phpdijet-xgam}
\end{figure}

Figure~\ref{fig:phpdijet-xgam}-left shows the dijet cross section in
diffractive photoproduction measured by ZEUS as a function of $x_\gamma^{jets}$
\cite{zeusdijet-gp},
the longitudinal momentum fraction of the parton that participated in
the hard scattering (see Fig.~\ref{fig:phpdijet-xgam}-right diagram), 
reconstructed from the dijet momenta.
Resolved events dominate in the low-$x_\gamma^{jets}$ region
while the direct process is concentrated at $x_\gamma^{jets}$ close
to one. 
The cross sections compared to the NLO QCD prediction, exhibits a
factorisation breaking both in the direct and in the resolved parts. 
The NLO QCD prediction required a global factor 
of 0.34 to be able to describe the data.
Similar results have been measured by H1~\cite{h1dijet}.
\\

{\bf Summary of factorisation tests}\\
 In its validity domain ($Q^2 >> 1 \rm\ GeV^2$ in $ep$ collisions), the
QCD factorisation holds, as confirmed by charm and dijet productions.
In $p\bar{p}$ collisions, at Tevatron, a breakdown of the QCD factorisation
of a factor 5 to 10 is observed for one gap formation and no (or weak) 
additional gap suppression is observed for a second gap formation.

\section{RATIO OF DIFFRACTIVE TO INCLUSIVE CROSS SECTIONS} 

\begin{figure}[h]
 \begin{center}
  \vspace*{-0.5cm}
  \epsfig{figure=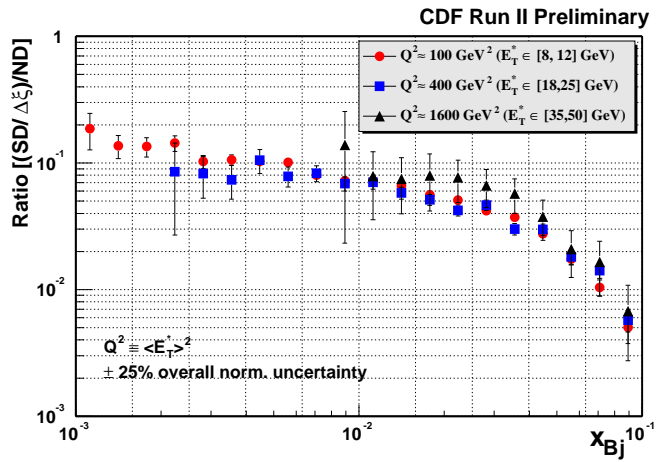,width=0.55\textwidth}
 \end{center}
 \vspace*{-1.3cm}
\caption[]{Ratio of single to non-diffractive cross
sections in $p\bar{p}$ collisions.} 
\label{fig:CDF_II}
\end{figure}

Figure~\ref{fig:CDF_II} presents CDF results from Tevatron Run II on the 
ratio between single-diffractive to non-diffractive cross 
sections~\cite{CDF_II}.
In the present case the scale $Q^2$ corresponds to $E^2_T$, where $E_T$
is the average transverse energy of the two jets. The ratio is observed 
to be independent of the $Q^2$ value.
\\

\begin{wrapfigure}[30]{L}{0.65\linewidth}
 \vspace*{-1.0cm}
 \epsfig{figure=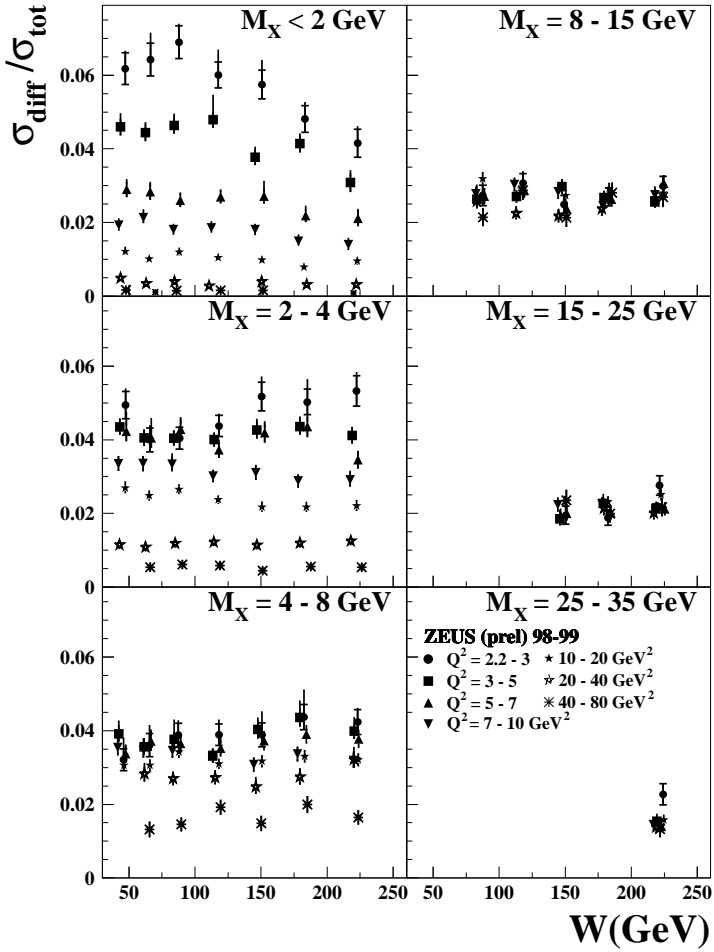,width=0.62\textwidth}
\caption{Ratio of single to non-diffractive cross
sections in $ep$ collisions.} 
 \vspace*{-3.0cm}
\label{fig:ratio}
\end{wrapfigure}

The ratio of the diffractive to inclusive cross sections in $ep$
collisions, as measured by ZEUS \cite{zeusratio}, 
is presented on Fig.~\ref{fig:ratio},
as a function of $W$, for different values of \qsq\ in different
intervals of the invariant hadronic final state mass, $M_X$. 
Analysing this ratio, three different behaviors can be pointed out.
i) For $M_X > 2$ GeV the ratio is flat in $W$, i.e.\ inclusive
diffraction has the same energy dependence than the total cross section
in DIS,
which is not consistent with a simple two gluon exchange. Indeed, in a
two gluon exchange the cross section is expected to grow like the gluon
density squared, while the total cross section is proportional to the 
gluon density, giving a ratio proportional to the gluon density, i.e.
increasing fast for decreasing \xbj\ values (or for increasing $W$ at
fixed \qsq). 
\\

ii) For $M_X > 8$ GeV, no $Q^2$ dependence is observed, i.e.\ 
the diffractive exchange evolves with $Q^2$ just like the inclusive
cross section, according to the DGLAP evolution. This can be understood
intuitively as when $M_X$ is large, \bet\ is small 
($\beta \simeq Q^2/(Q^2+M^2_X)$) and the virtual
photon "sees" only a small fraction of the diffractive exchange
momentum. This offers a large kinematic space to radiate gluons,
similarly to the "one" gluon exchange of the boson gluon fusion (main) 
responsible of the 
total cross section in DIS. In other terms, in this kinematic region the photon
cannot distinguish between DIS and a diffractive
exchange\\
iii) For $M_X < 2$ GeV. When $M_X$ decreases, \bet\ increases and the
photon "sees" more and more of the diffractive exchange up to the limit
of $\beta \rightarrow 1$ where the full diffractive exchanged momentum 
is connected to the photon and no gluon radiation is possible. 
In this case we can expect that the cross
section behaves like the gluon density squared. The large \bet\
region is dominated by vector meson production. Very interesting results
\cite{h1jpsi,zeusjpsi} show that, indeed, the $J/\Psi$ exclusive production 
can be described by QCD models based on 2 gluon exchange.

\section{CONCLUDING REMARKS} 
After almost ten years of research at HERA and at the
Tevatron, we are reaching a perturbative QCD understanding of hard
diffraction. The partonic structure of the diffractive exchange has been
measured and is found to be dominated by gluons. The factorisation in terms of
structure functions holds in $ep$ collisions at large \qsq\ but
rescattering corrections are important in $p\bar{p}$ and in $ep$
collisions in photoproduction. A global
understanding of inclusive and exclusive hard diffractions has progressed
recently. Many more results and a deeper understanding are expected with
the coming data at HERA II, Run II at Tevatron, Compass and in a further
future at LHC. 

\section{ACKNOWLEDGMENTS} 
It is a pleasure to thank the organizers for the invitation and the 
perfect organization of this very interesting conference. I am also 
grateful to Alessia Bruni and Laurent Schoffel for valuable help and
discussions.
This work is supported by the Fonds National de la Recherche
Scientifique Belge (FNRS).

\end{document}